\documentclass[twoside,fleqn]{article} 
\usepackage{espcrc2}
\usepackage[english]{babel}
\newcommand{\ket}[1]{|#1\rangle}
\newcommand{\bra}[1]{\langle #1|}
\newcommand{\mean}[1]{\langle #1\rangle}
\newcommand{\order}[1]{{\mathcal O}(#1)}
\newcommand{\re}{{\rm Re\,}}
\newcommand{\ope}[1]{\widehat{#1}}
\newcommand{\tihop}{\ope{\rho}}
\newcommand{\A}{\ope{A}}
\newcommand{\nham}{\ope{h}}
\newcommand{\canmean}[1]{\mean{#1}^{{\rm can}}}
\newcommand{\tr}{{\rm Tr}\,}
\newcommand{\ham}{\ope{H}}
\newcommand{\e}{{\rm e}}
\newcommand{\dd}{{\rm d}}
\newcommand{\half}{\frac{1}{2}}
\newcommand{\ci}{{\rm i}}
\newcommand{\vep}{\varepsilon}

\title{Using lattice methods in non-canonical quantum statistics}

\author{J.~Lukkarinen\address{Helsinki Institute of Physics,
  P.O.~Box 9, 00014 University of Helsinki, Finland \\
  electronic mail: {\tt jani.lukkarinen@helsinki.fi}}%
  \thanks{Presented as a poster at Lattice '99, Pisa, Italy.}}

\begin{document} 

\begin{abstract}
We define a natural coarse-graining procedure which can be applied to
any closed equilibrium quantum system described by a density matrix
ensemble and we show how the coarse-graining leads to the Gaussian and
canonical ensembles.  After this motivation, we present two ways
of evaluating the Gaussian expectation values with lattice
simulations. The first one is computationally demanding but general,
whereas the second employs only canonical expectation values but it is
applicable only for systems which are almost thermodynamical.
\end{abstract}

\maketitle

\section{Introduction}

The usefulness of the canonical ensemble in statistical mechanics is
remarkable.  The standard explanation of this success relies in taking
the ``thermodynamical limit'' which corresponds to increasing the
volume of the system to infinity while keeping all  the relevant
intensive quantities, i.e.\ densities, fixed and finite.  From this
point of view, the canonical ensemble should not have much utility for
small systems consisting of only a few particles.  This, however, does
not seem to be the case, and in the following we propose a new
approach which explains why and how the canonical ensemble can help
also in the analysis of small equilibrium systems.  We also explain
how lattice  simulations in general can be employed in this analysis.

\section{Ensembles from coarse-graining of the
energy fluctuation spectrum}

The standard approach to quantum statistics \cite{Balian1} uses a
density matrix $\tihop$, which is a non-negative,  hermitian,
trace class operator normalized to one and which  gives the
expectation value of an observable $\A$ by the formula
$\mean{\A}=\tr(\A\tihop)$.  In some complete eigenbasis
$\ket{\psi_i}$, the density matrix can thus be expanded as
$\tihop=\sum_i p_i \ket{\psi_i}\bra{\psi_i}$, where the eigenvalues
$p_i$ satisfy $p_i\ge 0$ and the normalization condition 
$\sum_i p_i = 1$.  

Suppose now that the system has a discrete energy spectrum, which in
quantum mechanics is achieved for every potential that grows
sufficiently fast at infinity. 
An equilibrium, i.e.\ time-independent, ensemble is then
given by a density matrix which has time-independent eigenvalues and
which satisfies $[\ham,\tihop]=0$. In this case, the eigenvectors
$\psi_i$ can be chosen so that they are also energy eigenvectors with
an eigenvalue $E_i$.

After these preliminaries, it is not hard to see that for {\em any}\/
equilibrium ensemble which is given by a density matrix and which has
energy as the only relevant conserved quantity,  we can find a smooth
mapping $F$ so that $\tihop=F(\ham)$, i.e.\ that  $\tihop = \sum_i
F(E_i) \ket{\psi_i}\bra{\psi_i}$.  We will call such a smooth mapping
the {\em fluctuation spectrum}\/ of the ensemble and we will use the
term {\em precanonical ensemble}\/ for those equilibrium ensembles
which satisfy:
\begin{enumerate}
\item The canonical partition function converges, $\tr \e^{-\beta\ham}
<\infty$, for all $\beta>0$.
\item The energy fluctuations decay at least exponentially at high
energies: $\e^{\beta E} F(E)$ is a rapidly decreasing function
\cite{RCA} for all $\beta<\beta_+$, where $\beta_+>0$ is a parameter.
\end{enumerate}

The following representation is then valid for any precanonical
ensemble  and for all $0<\beta<\beta_+$  as long as $\A$ is a
positive
observable which satisfies $\tr[\A\e^{-\beta\ham}] <\infty$ in the
same range of $\beta$:
\begin{equation}\label{e:trAcan}
\tr\!\bigl[\A F(\ham)\bigr] =  \int_{\beta-\ci
 \infty}^{\beta+\ci\infty} \!\!  {\dd w\over 2\pi\ci} \bar{F}(w)
 \tr\!\bigl[\A\e^{-w \ham}\bigr],
\end{equation}
where $\bar{F}$ is the Laplace transform of $F$ and the integrand in
the above equation is an analytic function in the half-plane  $0<\re
w<\beta_+$. This result follows from Fourier-transform  formulae for
rapidly decreasing functions; the precise mathematical details  can
be found from \cite{JML:tsallis}.

The value of the integral representation (\ref{e:trAcan}) can be
computed by saddle point methods and, for instance when $\A=\ope{1}$,
there is a unique positive saddle point  $\beta$ which dominates the
value of the integral.   For $\A=\ope{1}$ it can be solved for small
$\beta$ from the saddle point equation 
\begin{equation}\label{e:sadp}
\canmean{\ham}_\beta = E + \beta\vep^2 + \order{\beta^2},
\end{equation}
where, assuming the normalization  $\int\! F=1$,
\begin{equation}\label{e:Fpar}
E = \int\!\! \dd x F(x) x,\quad \vep^2 = \int\!\!\dd x F(x) (x-E)^2.
\end{equation}

Since the precise form of the fluctuation spectrum is difficult to
measure, we need some way of parameterizing its large scale
properties.  As usual, these can be extracted from the original
fluctuation spectrum by a coarse-graining transformation---here we
used a convolution with a Gaussian distribution,
\[
F(x) \mapsto F_\Lambda(x) \equiv  \int_{-\infty}^\infty \!\!  \dd y
F(y) {1\over \sqrt{2 \pi \Lambda^2}}  \e^{-{1\over 2 \Lambda^2}
(x-y)^2}.
\]
Under this transformation, the Laplace-transform in (\ref{e:trAcan})
will change to  $\bar{F}_\Lambda(w) = \e^{\half\Lambda^2 w^2}
\bar{F}(w)$.

When $\Lambda$ approaches infinity it is clear that the positive
saddle point value $\beta$ must go to zero.  Since $\bar{F}_\Lambda$
is analytic near the origin, we can in this limit use the approximation 
$E w + \half \vep^2 w^2$ for $\ln\bar{F}_\Lambda(w)$---the parameters 
are obtained from (\ref{e:Fpar}) by replacing $F$ with $F_\Lambda$.
Taking the inverse Laplace-transform then shows that this corresponds
to using the Gaussian ansatz $F_\Lambda(x) = G_{\vep}(E-x)$ for the
fluctuation spectrum.  This should not come as a surprise; the
argumentation is the same as used with the central limit theorem of
probability theory. 

For large $\Lambda$ the positive saddle point typically becomes
dominant. On the other hand, the trace left in the positive saddle
point  approximation is simply the canonical trace,
$\tr(\A\e^{-\beta\ham})$, and often the canonical expectation value
becomes a good approximation of the coarse-grained one.  The precise
condition for the use of the canonical ensemble can be given in terms
of the canonical variance
$\sigma^2=\canmean{(\ham-\mean{\ham})^2}_\beta$ and the normalized
canonical energy operator $\nham = (\ham-\mean{\ham})/\sigma$.  The
condition for using the canonical approximation  for the partition
function and the one for using the canonical expectation value
for a positive observable $\A$ are, respectively,
\[
a \equiv { \sigma^2\over 2 \vep^2} \ll 1\quad{\rm and}\quad
a {\mean{\A\nham^2}/ \mean{\A}} \ll 1.
\]
Similarly, the Gaussian ensemble can be used if the left hand sides
in the previous equations are not too large.  A more complete
explanation of these results can be found in \cite{JML:tsallis}.

The canonical approximation of the Gaussian expectation values has
already been analysed in \cite{JML:gauss} and we quote here only the
results:  The simple bounds, already referred to in the above,  for
the approximation of positive observables are
\[
-a { \mean{\A\nham^2}\over \mean{\A} } \le 
 \ln {\mean{\A}^{\rm gauss}_{E_,\vep}\over  
 \mean{\A}^{\rm can}_\beta } \le a
\]
and this approximation can be improved for $a\approx 1$ 
by using the asymptotic series
\begin{equation}\label{e:asser}
\mean{\A}^{\rm gauss}_{E,\vep} = \mean{\A}^{\rm can}_\beta + 
  \mean{(1-\nham^2)\A} a + \order{a^2}.
\end{equation}

\section{Gaussian ensemble on a lattice}

There are two different ways of using lattice simulations in the
evaluation of Gaussian expectation values.  In the direct approach,
the lattice approximation is applied to the complex temperature trace
in (\ref{e:trAcan}) which, after an exchange of the order of the
integration  and the continuum limit, leads to an integral kernel for
the lattice simulations.  Unfortunately the kernel is not a positive
function and the results are in most cases obtained as a delicate
cancellation of oscillations of the kernel. This, however, is likely
to be an unavoidable feature of any space-lattice simulation of
microcanonical expectation values as the energy wavefunctions
themselves are typically highly oscillatory.

This approach is most useful when canonical simulations  at a complex
temperature are possible.  Then the trace in the the integrand in
(\ref{e:trAcan})  can be evaluated in a number of points and the
integral computed by  a discrete Fourier-transform.  This would yield
results for a range of values of the parameters $E$ and $\vep$ and,
therefore, it would enable an inspection of a whole energy range at the
same time. The main difficulty in this approach  is, of course, in the
complex temperature lattice simulation with its oscillatory kernel
function.

If an expectation value at only one value of the parameters $E$ and
$\vep$ is needed, then a second alternative is also possible: perform
first the Fourier-transformation of the canonical lattice kernel and
do the lattice simulations with the resulting kernel function.  In
this case, the evaluation of the kernel function becomes an obstacle
slowing down the simulation.

The second approach to the Gaussian evaluation problem is to use the
asymptotic series given in (\ref{e:asser}) which requires the
computation of {\em canonical}\/ expectation values only.  The problem
in this case is to find the correct lattice operators which would
correspond to the different powers of $\nham$ in the continuum
limit.  We will now show how these can be found in a simple quantum
mechanical case and comment on some general features which should be
relevant also for field theory lattice simulations---a  more complete
analysis of this kind of lattice system can be found from
\cite{JML:latmom}.

Consider, for simplicity, a non-relativistic particle in a potential
$V(x)$.  The Hamiltonian of this system is $\ham={1\over 2 m}
\ope{p}^2 + V(\ope{x})$ and if the potential is bounded from below and
increases sufficiently fast at infinity, the complex temperature trace has 
a rigorous lattice approximation given by
\[
\tr \e^{-w \ham} = \lim_{L\to\infty} \int\!\! \dd^L x [L/(2\pi
w)]^{L/2} \e^{-{1\over w}P_L-w V_L},
\]
\vspace*{-3ex}
\[
P_L = {L\over 2 m}\sum_{k=1}^L\left|x_{k-1}-x_k\right|^2,\  V_L =
{1\over L} \sum_{k=1}^L V(x_k).
\]

\begin{table}
\caption{Lattice energy operators in terms of $M=L/2$,
$P_\beta={1\over \beta}P_L$ and $V_\beta=\beta V_L$.}
\label{t:latenop}
\begin{tabular}{l@{ = }ll@{ = }l} \hline 
\multicolumn{4}{c}{ }\\[-2.4ex]  ``$\beta H$'' & $c_1$ & $c_1$ & $  M
- P_\beta + V_\beta$ \\ ``$\beta^2 H^2$'' & $c_1^2+c_2$ & $c_2$ & $ M-
2 P_\beta$ \\ ``$\beta^3 H^3$'' & $c_1^3 + 3 c_1 c_2 + c_3$ & $c_3$ &
$2 (M- 3 P_\beta)$  \\[.4ex] \hline
\end{tabular}
\end{table}

A straightforward differentiation of this result then gives the
operators which will measure in the continuum limit the expectation
values of different powers of the Hamiltonian.  We have given the
first few of them in Table \ref{t:latenop}.  Two features of these
results are worth pointing out: first, the kinetic energy is given by
the operator $\half L-{1\over \beta}P_L$, which shows that $P_L$
diverges as the lattice size in the continuum limit and thus needs to
be ``renormalized''.  This reflects the well-known result that the
continuum path-integral is concentrated on paths which are continuous,
but non-differentiable.  Secondly, each power of the Hamiltonian needs
a separate renormalization term in the sense that using the powers of
the operator giving the expectation value, $c_1$, is not enough.  This
is exactly analogous to the situation of composite operators in field
theory.

\section{Conclusions}

We have introduced the Gaussian ensemble as a means of refining the
accuracy of the canonical ensemble and we have shown by a
coarse-graining procedure why this would have applications also for
non-thermal equilibrium systems.  The canonical, complex temperature,
lattice simulations offer one way of inspecting the behavior of the
Gaussian  expectation values.  A second way, applicable for systems
near the thermodynamical limit, uses correction terms which can be
computed with the well-established methods of canonical lattice
simulations.

\end{document}